\documentclass[%
 reprint,
superscriptaddress,
 amsmath,amssymb,
 aps,
prd,
]{revtex4-2}

\usepackage{graphicx}
\usepackage{bm}
\usepackage{amsmath}
\usepackage{mathtools}
\usepackage{comment}
\usepackage{xspace}

\usepackage{savesym}
\savesymbol{tablenum}
\usepackage{siunitx}
\restoresymbol{SIX}{tablenum}

\usepackage{xcolor}

\newcommand{\rmd}{\mathrm{d}}
\newcommand{\ali}{\mathrm{ali}}
\newcommand{\iso}{\mathrm{iso}}
\newcommand{\Hali}{\mathcal{H}_\ali}
\newcommand{\Hiso}{\mathcal{H}_\iso}
\newcommand{\SDR}[1][\vartheta]{\ensuremath{\mathrm{SDR}^\iso_{\ali, #1}}\xspace}
\newcommand{\chip}{\chi_\mathrm{p}}
\newcommand{\chieff}{\chi_\mathrm{eff}}
\newcommand{\ctls}{\cos\theta_{LS}}
\DeclarePairedDelimiterX{\infdivx}[2]{(}{)}{%
  #1\;\delimsize\|\;#2%
}
\newcommand{\DKL}{D_\mathrm{KL}\infdivx}

\usepackage[colorlinks=true, allcolors=darkblue]{hyperref} 
\definecolor{darkblue}{rgb}{0,0.2,0.6}
\usepackage{orcidlink}

\allowdisplaybreaks

\begin{document}

\title{Measurement of spin--orbit misalignment in binary black holes via the total spin}
\author{Charles F.\ A.\ Gibson\orcidlink{0009-0003-8690-8297}}
\email{charlesgibson2031@u.northwestern.edu}
\affiliation{Department of Physics \& Astronomy, Northwestern University, Evanston, IL 60208, USA}
\affiliation{Center for Interdisciplinary Exploration \& Research in Astrophysics (CIERA), Northwestern University, Evanston, IL 60201, USA}

\author{Isha Anantpurkar\orcidlink{0000-0002-5814-4109}}
\affiliation{Department of Physics, University of California at Santa Barbara, Santa Barbara, CA 93106, USA}

\author{Javier Roulet\orcidlink{0000-0003-3268-4796}}
\affiliation{TAPIR, Walter Burke Institute for Theoretical Physics, California Institute of Technology, Pasadena, CA 91125, USA}
\affiliation{Kavli Institute for Cosmological Physics, The University of Chicago, 5640 South Ellis Avenue, Chicago, Illinois 60637, USA}
\affiliation{School of Natural Sciences, Institute for Advanced Study, 1 Einstein Drive, Princeton, NJ 08540, USA}

\begin{abstract}
The degree of spin--orbit alignment in merging binary black holes is a powerful probe of their formation history.
However, assessing spin--orbit misalignment in individual events remains challenging, as most parameters commonly used to characterize precession tend to be poorly constrained.
In this work, we introduce the angle between the total spin and orbital angular momentum, $\theta_{LS}$, as an alternative measure of spin--orbit misalignment.
Using synthetic observations, we show that $\theta_{LS}$ retains more information to discriminate between the aligned- and isotropic-spins hypotheses than commonly used alternatives, including $\chi_{\rm p}$.
We then study this parameter on binary black hole mergers observed in the GWTC-5.0 catalog, and identify multiple events that are inconsistent with having aligned spins.
In several of the examples, this is not apparent from the posteriors of the effective spin parameters ($\chip$ or $\chi_\mathrm{eff}$) nor the individual spin--orbit tilts ($\theta_1$ or $\theta_2$) alone, and thus had not been previously identified.
We highlight GW241127\_061008 as a compelling case for a second-generation merger, with a large, tilted primary spin, unequal masses, and a large primary mass near the pair-instability-supernova mass gap.
Finally, we perform population inference using $\cos\theta_{LS}$ and confirm that the binary black hole population is inconsistent with either a purely isotropic or a purely aligned-spin distribution, and requires substantial spin--orbit misalignment.
\end{abstract}

\maketitle

\section{Introduction}

Binary black hole systems can be formed through a variety of physical processes, and through the detection of binary black hole mergers by the LIGO--Virgo--KAGRA (LVK) gravitational-wave detectors, we can gain insight into the formation channel of these systems \cite[e.g.,][]{MandelFarmer22}.
For example, tight compact-object binary systems can be formed from isolated stellar binaries in the galactic field, either assisted by stable \citep{Inayoshi2017, vandenHeuvel2017, Pavlovskii2017, GallegosGarcia2021, vanSon2022} or unstable \citep{Belczynski2007, Postnov2014, Klencki2021, Marchant2021, Olejak2021} mass transfer, or avoiding mass transfer altogether through a chemically homogeneous evolution that prevents stellar expansion
\citep{deMink2016, Mandel2016, Marchant2016}.
Alternatively, these systems could be assembled and hardened dynamically in dense environments such as globular \citep{PortegiesZwart2000, Rodriguez2015}, nuclear \citep{Antonini2016, Petrovich2017}, or young open clusters \citep{Ziosi2014} through three- (or more) body interactions.
Another possibility involves Kozai--Lidov oscillations \citep{Lidov1962,Kozai1962} in hierarchical triples \citep{Liu2015, Silsbee2017,
Antonini2018, Antonini2017, VignaGomez2021} (or higher-multiplicity systems \citep{Hamers2019}), which can excite the orbital eccentricity and enhance gravitational wave emission.
Binary black hole systems can also form in active galactic nuclei disks \citep{McKernan2012, Bartos2017, Tagawa2026}, driven by interactions with gas and stars therein.

Through the observation of the masses, spin magnitudes, orientations, eccentricity, and merger rate evolution of binary black hole mergers, we can gain hints into the formation channels of these individual binary black hole systems. In this work, we focus on the spin orientations.
For binaries formed in isolation, spins are expected to be co-aligned with the orbit through mass transfer or tides, although supernova kicks at black hole formation can misalign the orbit \citep{Kalogera2000,Gerosa2018,Steinle2021}, and stable mass transfer can misalign the spin of the donor \citep{Stegmann2021}.
Conversely, dynamically assembled binaries generally have isotropically distributed spins \citep{PortegiesZwart2000, Rodriguez2015}.
This said, a small fraction of dynamically assembled systems may potentially be aligned by gas accretion of disrupted stars \citep{Kiroglu+25}.
Furthermore, binary black hole mergers in globular \citep{Hong+18,Arca_sedda+24,OConnor+26} and open \citep{DiCarlo+19,Banerjee+23} cluster environments can exhibit characteristics of both isolated binary evolution and traditional dynamical formation.
In the case of primordial spin-orbit alignment of binaries in clusters, a single, strong interaction will not necessarily yield an isotropic distribution of spin-orbit orientation \citep{Tranni+21}, requiring several successive dynamical encounters to isotropize the binary \citep{Martinez+26}.
Finally, triple systems can produce spins preferentially perpendicular to the orbit due to the Kozai--Lidov mechanism \citep{Antonini2018, Rodriguez2018, Liu2018}.

As a key discriminator between formation channels, the degree of spin--orbit alignment is a subject of intensive research.
Multiple spin orientation parameters, both physically and observationally motivated, have been studied.
Perhaps the most widely used are the effective aligned spin
\begin{equation}
    \chieff = \frac{m_1\bm{\chi}_1 + m_2 \bm{\chi}_2}{m_1 + m_2} \cdot \bm{\hat L}
\end{equation}
and effective precession spin
\begin{equation}
     \chip = \max\left\{\chi_{1\perp}, \frac{3 + 4q}{4 + 3q}\,q\,\chi_{2\perp}\right\}
\end{equation}
(where $\bm{\chi}_1, \bm{\chi}_2$ are the dimensionless spins of the two objects, $\chi_{1\perp}, \chi_{2\perp}$ their in-plane magnitudes, and $q=m_2/m_1$).
$\chieff$ is relatively well measured and conserved to second post-Newtonian order \cite{Racine08}.
Its sign can be used distinguish aligned from anti-aligned spin--orbit configurations, but otherwise $\chieff$ contains a limited amount of information about the spin orientations, as the in-plane spin components are discarded.
$\chip$ was introduced to model the average rate of precession at leading post-Newtonian order \citep{SchmidtOhmeHannam+15}, but has some shortcomings for the purpose of assessing spin--orbit misalignment in individual events: posterior distributions of $\chip$ are often broad, and most commonly adopted prior probability densities vanish at $\chip=0$, which corresponds to the astrophysically interesting case of aligned-spin configurations.
These features complicate both the interpretation of individual events and the construction of statistical tests for spin alignment.
For these and other reasons, several studies have also considered modified $\chip$ definitions \citep{Gerosa2021, Thomas2021}; the precessing signal-to-noise ratio $\rho_{\rm p}$~\citep{Fairhurst2020}; a taxonomy of phenomenological parameters \citep{Gangardt2021}; the spin azimuths about the orbital angular momentum (relative to the orbital separation or to each other) \citep{Varma2021}; or the total spin azimuth about the total angular momentum relative to the line of sight \citep{Roulet2022}.
The individual black hole tilts ($\theta_1, \theta_2$) are also commonly used to quantify spin--orbit misalignment, with $\cos\theta_{1,2} = \bm{\hat L} \cdot \bm{\hat \chi}_{1,2}$. 
However, these parameters are often poorly constrained. Accurately measuring their true distribution is difficult at the current catalog size and uncertain even with a catalog of $\sim1500$ events \citep{VitaleMould2025}, further motivating the need for a better measured spin orientation parameter.

In this work, we argue that the tilt angle $\theta_{LS}$ between the total spin and the orbital angular momentum constitutes a sensitive and interpretable diagnostic of spin--orbit misalignment in individual binary black hole systems.
We begin in Section~\ref{sec:methods} by motivating four candidate parameters to measure spin--orbit alignment, along with developing the statistical formalism and testing on synthetic events which parameter is the most informative.
In Section~\ref{sec:LVK}, we use $\ctls$ to analyze the spin--orbit alignment of individual events in GWTC-5.0.
In Section~\ref{sec:population}, we construct a series of models for the astrophysical distribution of $\ctls$ and find that all agree on the existence of spin--orbit misalignment in the astrophysical population.
We conclude in Section~\ref{sec:discussion}.

\section{Identifying informative spin-alignment parameters}\label{sec:methods}

While, in principle, all the astrophysical information from a signal is contained in the full ($\sim 15$D) posterior distribution, it is desirable to identify the most informative combinations of parameters.
This aids interpretability, intuition, and visualization of results. Moreover, since lower-dimensional distributions are generally more tractable to density estimation or importance sampling methods, approximations (such as neglecting uninformative parameters) can be designed to improve numerical convergence.

In this section, we seek a single parameter that best encapsulates the information about spin alignment.
Formally, we aim to identify a parameter involving the spin orientations whose marginal posterior distribution retains the greatest statistical power to distinguish between an astrophysical model in which binaries have spins aligned with the orbit, $\Hali$, versus isotropically distributed, $\Hiso$ (as predicted by the most simplistic models of isolated or dynamical formation, respectively).
In Section~\ref{sec:param_candidates} we motivate the angles $\theta_{LS}$ and $\beta$ as candidate spin-misalignment parameters alternative to $\chip$ and $\cos\theta_1$, in Section~\ref{sec:formalism} we formalize our criterion of a maximally informative parameter, and in Section~\ref{sec:performing_parameter_test} we assess the information content of the three parameters using synthetic gravitational wave data.

\subsection{Candidate spin-misalignment parameters}\label{sec:param_candidates}

\begin{figure}
    \centering
    \includegraphics[width=.75\linewidth]{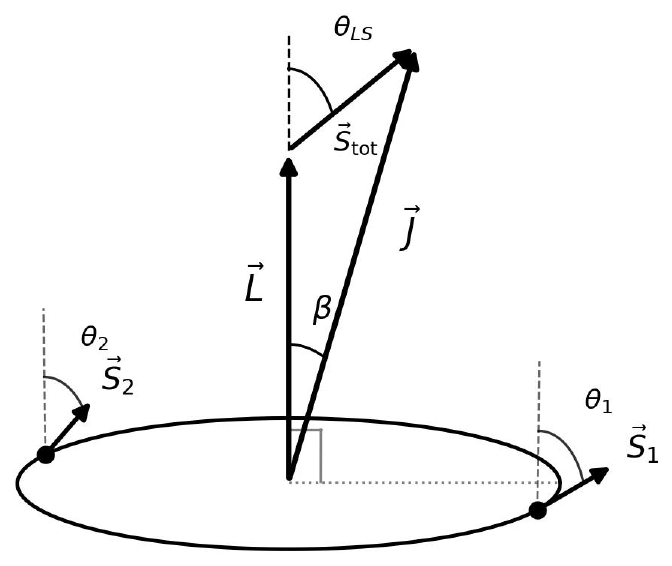}
    \caption{The geometry of a binary black hole system.
    The spins of each black hole are denoted by $\bm{S}_i$ (with the total spin $\bm{S}_{\rm tot}=\bm{S}_1+\bm{S}_2$), the orbital angular momentum is expressed as $\bm{L}$, and the total angular momentum is $\bm{J} = \bm{L}+\bm{S}_\mathrm{tot}$.
    $\theta_{LS}$ is the angle between $\bm{L}$ and $\bm{S}_{\rm tot}$.
    $\beta$ is the angle between $\bm{J}$ and $\bm{L}$. $\theta_1$ and $\theta_2$ are the angles of the individual spin tilts.}
    \label{fig:bbh_geometry}
\end{figure}

Using the geometry of the binary black hole merger outlined in Fig.~\ref{fig:bbh_geometry}, we select two alternative parameters to compare with $\chi_{\rm p}$ and $\cos\theta_1$: $\theta_{LS}$ and $\beta$. The first parameter, $\theta_{LS}$, is defined as the angle between the orbital angular momentum $\bm{L}$ and the total spin $\bm{S}_{\rm tot} = \bm{S}_1 + \bm{S}_2$:
\begin{equation}
\label{eq:costhetals}
    \cos{\theta_{LS}}
    = \bm{\hat{S}}_\mathrm{tot}\cdot\bm{\hat{L}} = \frac{
        m_1^2\bm{\chi}_1 + m_2^2\bm{\chi}_2
    }{
        \left|m_1^2\bm{\chi}_1 + m_2^2\bm{\chi}_2\right|
    } \cdot \bm{\hat{L}}.
\end{equation}
$\ctls$ has several compelling features compared to $\chi_\textrm{p}$:
First, it follows a simple distribution under either the aligned ($\ctls = 1$) or isotropic ($\ctls$ uniform in $(-1, 1)$) spin models.
In contrast, the $\chip$ distribution for the isotropic model is more complicated and depends on the spin magnitudes \citep{Iwaya2025}.
Second, its prior probability density does not vanish at the value associated with aligned spins, $\pi(\ctls{=}1) \neq 0$, while $\pi(\chip{=}0)=0$, which in practice can affect the convergence of population studies \citep{Kobayashi2026} and obscure the interpretation of individual events.
Third, it can distinguish a configuration in which the spins are aligned with the orbit ($\ctls=1$) from one in which they are anti-aligned ($\ctls=-1$); both configurations correspond to $\chip=0$ but have very different astrophysical implications.
On the flip side, $\ctls$ becomes ill defined when the spin is small while $\chip$ does not.
Also, $\ctls=1$ does not necessarily mean that the spins are aligned, but rather that the in-plane spins of the two objects cancel.
Compared to the individual tilt angles $\theta_1$ and $\theta_2$, we expect the total-spin tilt $\theta_{LS}$ to be more sensitive to signatures of precession in the data: even when precession is detected, the data may only indicate that at least one spin is misaligned, while still allowing either the primary or secondary spin to be individually aligned, particularly in comparable-mass systems.
In Section~\ref{sec:LVK} we will show that these situations do arise in real-data examples.

The second angle, $\beta$, is the angle between the orbital angular momentum and the total angular momentum: $\cos\beta = \bm{\hat{L}}\cdot\bm{\hat{J}}$. This angle is promising because it impacts the magnitude of the modulations and phase evolution of the waveform in the early inspiral \citep{Fairhurst+20}. Particularly, the parameter $b=\tan{\beta/2}$ is used to directly compute the waveform.
A nuisance is that, under the isotropic spin hypothesis, the distribution for $\beta$ is not isotropic, and depends on the spin magnitudes and the reference frequency at which the system is specified.
Conversely, although $\chi_{\rm p}$ and $\cos{\theta_{LS}}$ also evolve with the frequency (for misaligned systems), their distributions are constant for both the aligned and the isotropic spin model (an aligned-spin distribution at early times remains aligned, and isotropic remains isotropic \citep{Bogdanovic2007, Gerosa2014}).
We test $\ctls$, $\cos{\beta}$, $\chip$, and $\cos\theta_1$ using the formalism outlined in Section~\ref{sec:formalism}.

\subsection{Formalism}\label{sec:formalism}

We formalize the problem of finding the most informative parameter as follows:
By Neyman--Pearson lemma \cite{NeymanPearson33}, given data $d$ the most powerful statistic to distinguish the two hypotheses (isotropic vs.\ aligned spins) is the Bayes factor
\begin{equation}
    \label{eq:bayes_factor}
    \mathcal B^\iso_\ali(d)
    = \frac{
        p(d \mid \mathcal \Hiso)
    }{
        p(d \mid \mathcal \Hali)
    }
    = \frac{
        \int \rmd \bm\theta \, p(d\mid\bm\theta) \, \pi(\bm\theta \mid \Hiso)
    }{
        \int \rmd \bm\theta \, p(d\mid\bm\theta) \, \pi(\bm\theta \mid \Hali)
    }
    .
\end{equation}
Note that the models we will compare in this work only differ on the astrophysics---encoded in the prior $\pi(\theta \mid \mathcal H)$, $\mathcal H \in\{\Hali, \Hiso\}$---while the likelihood $p(d \mid \bm \theta)$ modeling the signal and detector is the same for both.
$\log \mathcal B^\iso_\ali(d)$ can be interpreted as the information in an observation $d$ in favor of $\Hiso$ over $\Hali$ \cite[p.~5]{Kullback1959}.
Its expected value is the Kullback--Leibler divergence; for example,
\begin{equation}
    \label{eq:dkl}
    \DKL{p(d \mid \Hiso)}{p(d \mid \Hali)} 
    = \int \rmd d \,
        p(d \mid \Hiso)
        \log \mathcal B^\iso_\ali
\end{equation}
is the mean discrimination information between these models if the true population follows $\Hiso$.

We will use these concepts to define a ``most informative'' parameter.
We single out one spin-misalignment parameter $\vartheta$ that takes a value $\vartheta_\ali$ under the aligned-spin model, so that $\bm\theta \equiv (\vartheta, \bm \theta')$ and $\pi(\vartheta \mid \Hali) = \delta(\vartheta - \vartheta_\ali)$.
For example, $\vartheta$ may represent $\chip$, and $\chi_{\mathrm{p}, \ali} = 0$.

From Bayes' theorem we obtain the identity
\begin{multline}
    p(d \mid \Hiso) \,
    p(\vartheta_\ali \mid d, \Hiso) \\
    = 
    \pi(\vartheta_\ali \mid \mathcal \Hiso) \,
    p(d \mid \vartheta_\ali, \mathcal \Hiso),
\end{multline}
which we use to rewrite Eq.~\eqref{eq:bayes_factor} as
\begin{equation}
    \label{eq:bayes_factor_2}
    \mathcal B^\iso_\ali(d)
    = \frac{
        \pi(\vartheta_\ali \mid \mathcal \Hiso)    
    }{
        p(\vartheta_\ali \mid d, \Hiso)
    }
    \cdot
    \frac{
        p(d \mid \vartheta_\ali, \mathcal \Hiso)
    }{
        p(d \mid \mathcal \Hali)
    }.
\end{equation}
We will make the approximation of neglecting the second factor in Eq.~\eqref{eq:bayes_factor_2}:
\begin{equation}
    \label{eq:approx}
    \begin{split}    
        \frac{
            p(d \mid \vartheta_\ali, \mathcal \Hiso)
        }{
            p(d \mid \mathcal \Hali)
        }
        &=
        \frac{
            \int \rmd \bm \theta' \,
            p(d \mid \vartheta_\ali, \bm \theta') \,
            \pi(\bm \theta' \mid \vartheta_\ali, \Hiso)
        }{
            \int \rmd \bm \theta' \,
            p(d \mid \vartheta_\ali, \bm \theta') \,
            \pi(\bm \theta' \mid \Hali)
        } \\
        &\approx 1,
    \end{split}
\end{equation}
which amounts to using the so-called Savage--Dickey ratio (SDR) of the marginal prior to the posterior on our selected parameter at $\vartheta_\ali$ to estimate the Bayes factor:
\begin{equation}
    \label{eq:bayes_factor_approx}
    \mathcal B^\iso_\ali(d)
    \approx \SDR(d)
    \coloneqq \frac{
        \pi(\vartheta_\ali \mid \mathcal \Hiso)    
    }{
        p(\vartheta_\ali \mid d, \Hiso)
    }.
\end{equation}
As desired, $\SDR(d)$ is an approximation to the optimal statistic $\mathcal B^\iso_\ali(d)$ which only depends on the 1D posterior for the selected parameter $\vartheta$.
Our goal is to identify which $\vartheta$ makes this approximation as good as possible; as discussed above, we will compare $\chip$, $\theta_1$, $\beta$, and $\theta_{LS}$.

To motivate and interpret this approximation, we note that the Savage--Dickey ratio exactly matches the Bayes factor when the discarded parameters are either unobservable or insensitive to the models we are comparing.
In the first case, if $\bm \theta'$ produce no observable effect, i.e., $p(d \mid \vartheta_\ali, \bm \theta') \approx p(d \mid \vartheta_\ali)$, the $\bm \theta'$ integrals are unity and the likelihoods cancel out.
In practice we expect this situation to apply approximately to some spin components that are difficult to measure.
In the second, if both models predict the same distribution for $\bm \theta'$, i.e., $\pi(\bm \theta' \mid \vartheta_\ali,\mathcal H_\iso) \approx \pi (\bm\theta' \mid \Hali)$, the models are said to be nested, and the numerator and denominator of Eq.~\eqref{eq:approx} coincide.
This case is relevant to extrinsic parameters and, for the purposes of this work, the masses of the black holes.

Motivated by Eqs.~\eqref{eq:dkl} and \eqref{eq:bayes_factor_approx}, we will define the most informative parameter as that which maximizes
\begin{multline}
    \label{eq:dkl_1par}
    \tilde D_\mathrm{KL, \vartheta}\infdivx{p(d \mid \Hiso)}{p(d \mid \Hali)} \\
    \coloneqq
    \int \rmd d \,
        p(d \mid \Hiso)
        \log \SDR,
\end{multline}
i.e., whose marginal posterior distribution, on average, contains the most information to discriminate an aligned-spin population from an isotropic-spin one.
We construct a Monte Carlo estimator for $\tilde D_\mathrm{KL, \vartheta}$ as follows:
we simulate a catalog of $N$ events from $\Hiso$, infer their associated posteriors, and use importance sampling to estimate
\begin{equation}
\begin{split}
    \tilde D_\mathrm{KL, \vartheta}\infdivx{p(d \mid \Hiso)}{p(d \mid \Hali)}
    &\approx
    \frac 1N \sum_{i=1}^N
    \log \SDR(d_i); \\
    d_i &\sim p(d \mid \Hiso). \label{eq:dkl_1par_approx}
\end{split}
\end{equation}
In turn, we compute each $\SDR(d_i)$ via Eq.~\eqref{eq:bayes_factor_approx}, using samples from the marginal prior and posterior for $\vartheta$ to estimate the corresponding densities, see Section~\ref{sec:performing_parameter_test} for details.

As a note, we use $\tilde D_\mathrm{KL, \vartheta}\infdivx{p(d \mid \Hiso)}{p(d \mid \Hali)}$ as the metric and not $\tilde D_\mathrm{KL, \vartheta}\infdivx{p(d \mid \Hali)}{p(d \mid \Hiso)}$, so that the simulated spins are actually misaligned, allowing us to study their richer phenomenology.

\subsection{Information content of spin-misalignment parameters}
\label{sec:performing_parameter_test}

We expect that the most informative parameters $\vartheta$ will have a low $\SDR$ value when the spin and the orbit are aligned, and a high $\SDR$ when they are misaligned.
Parameters that are not as strongly indicative of the precession will have little effect on the Savage--Dickey ratio.

We use $\approx3000$ posterior distributions for synthetic events from \citet{Roulet+24}, available at \citep{Roulet2024zenodo}, whose individual spin tilts are isotropically distributed and dimensionless spin magnitudes ($\chi\in[0, 1)$) are distributed as $p(\chi)=3\chi^2$ (``volumetric" spin prior).
This choice of prior aims to cover more-or-less homogeneously the space of possible waveform morphologies, although note that the astrophysical population has been observed to have smaller spin magnitudes \citep{Farr2017,Roulet2019, Miller2020, PopulationO3, Callister2024}. 
We impose a detectability cut on top of this prior, approximated by the criterion that the squared expected signal-to-noise ratio in the LVK detector network is $\langle h \mid h \rangle > 70$.

Expanding on this dataset, we consider an additional $\approx3\times10^4$ samples on which we do not perform parameter inference.
The purpose of these additional samples is to accurately estimate the effective prior, which results from the analytical astrophysical prior and the detectability criterion applied, and is used in computing the Savage--Dickey ratio. These $\approx3.3\times10^4$ samples were evenly distributed between three bins in detector-frame chirp-mass $\mathcal {M}$ (1--\SI{5}{M_\odot}, 5--\SI{25}{M_\odot}, 25--\SI{125}{M_\odot}), allowing us to investigate selection effects in mass. Within each bin, we adopted a uniform prior in component masses within $q \equiv m_2 / m_1 > 1/20$.
In all the injections, the parameters of the system were specified at a reference gravitational-wave frequency of $f_\mathrm{ref} = \SI{50}{\hertz}$.

Using the posterior distributions and the effective prior described above, we estimate $\SDR$ for each event and misalignment parameter $\vartheta~\in~\{\ctls, \cos\beta, \chip, \cos\theta_1\}$ using Eq.~\eqref{eq:bayes_factor_approx}. To do so, we construct a bin in $\vartheta$ containing $\vartheta_\ali$, with width $\Delta\vartheta$ chosen adaptively so that the bin contains at least a fraction $\epsilon = \num{2.5e-3}$ of both the prior and posterior probability mass, in order to keep sampling noise under control.
Because of this requirement, our procedure tends to underestimate the Savage--Dickey ratio for events where $\vartheta$ is constrained away from $\vartheta_\ali$ (in which case the bin $\Delta\vartheta$ gets enlarged until it contains appreciable probability and $p(\vartheta_\ali \mid d)$ is overestimated), so it should be interpreted as a conservative lower bound.
With the logic that the bin needs to extend at most up to $\vartheta_\mathrm{true}$ in order to contain a small fraction $\epsilon$ of the posterior, we can determine the maximum Savage--Dickey ratio obtainable through this estimator:
\begin{align}
    \SDR(d)
    &\approx \frac{
        \int_{\Delta\vartheta} \pi(\vartheta) \, \rmd \vartheta
    }{
        \int_{\Delta\vartheta} p(\vartheta \mid d) \, \rmd \vartheta
    } \\
    &\lesssim \frac{
        \int_{\vartheta_\mathrm{true}}^{\vartheta_\ali} \pi(\vartheta) \, \rmd \vartheta
    }{
        \epsilon
    }.\label{eq:max_bf}
\end{align}

\begin{figure*}
    \centering
    \includegraphics[width=\linewidth]{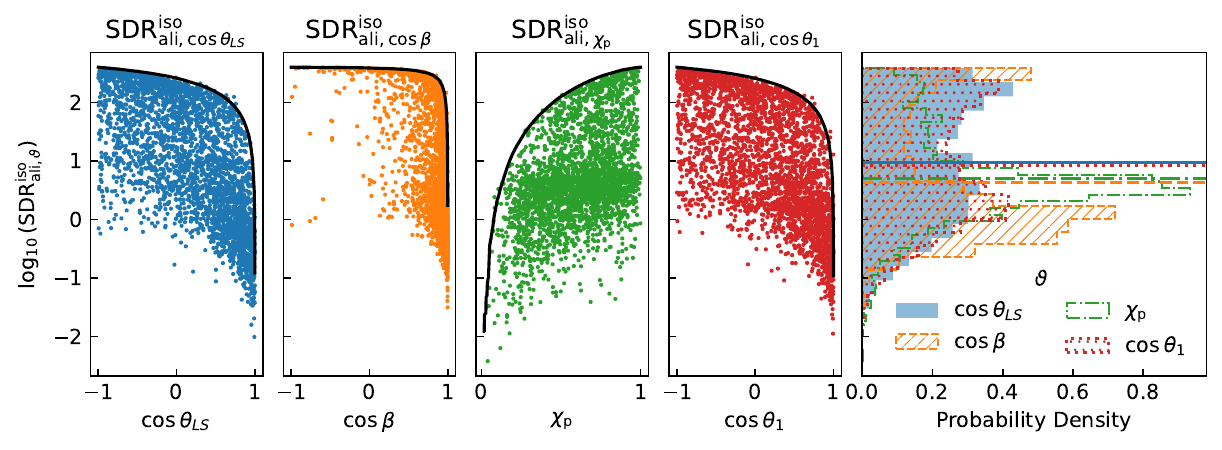}
    \caption{Savage--Dickey ratios \SDR (see Eq.~\eqref{eq:bayes_factor_approx}) for the four tested parameters $\vartheta \in \{\ctls, \cos\beta, \chip, \cos\theta_1\}$, for $\approx 3000$ synthetic events. In the first four panels, the horizontal axis corresponds to the true (unobservable) values of the injections while the vertical axis shows the corresponding (observable) $\SDR(d)$. The solid black curves provide an estimate of the maximum value of \SDR for each parameter that our computation algorithm allows (Eq.~\eqref{eq:max_bf} with $\epsilon = \num{2.5e-3}$); values near this line may be interpreted as lower bounds. The farthest right panel outlines the distribution of the Savage--Dickey ratios of each parameter. Horizontal lines denote the average $\log_{10}\SDR$, which correspond to the mean discrimination information $\tilde{D}_{\rm KL,\vartheta}$ between the isotropic- and aligned-spin models (Eqs.~\eqref{eq:dkl_1par} and \eqref{eq:dkl_1par_approx}).
    We find that $\cos{\theta_{LS}}$ has the highest discrimination information of the tested parameters. }
    \label{fig:savage-dickey}
\end{figure*}

These data are summarized in Fig.~\ref{fig:savage-dickey}.
In the first three panels we see that, as expected, each parameter shows a general trend of decreasing $\SDR$ as the parameter of each injected value approaches alignment ($\ctls=1$, $\cos\beta=1$, $\chip=0$, $\cos\theta_1=1$).
We also show the maximum Savage--Dickey ratio computable with our procedure as a function of the true parameter value (estimated from prior samples via Eq.~\eqref{eq:max_bf}), and find that it accounts very well for the lack of points above.
The right panel shows the histograms of $\log\SDR$ and their average, which by Eq.~\eqref{eq:dkl_1par_approx} is the Kullback--Leibler divergence or mean discrimination information in each parameter.
Across the combined chirp mass ranges, we can see that $\cos{\theta_{LS}}$ has the highest $\tilde D_\mathrm{KL, \vartheta}$, followed by $\cos\theta_1$, $\chi_{\rm p}$, and then $\cos{\beta}$, and thus we conclude that on average $\cos{\theta_{LS}}$ is the most informative parameter for spin--orbit misalignment out of those we tested.
The spin of the secondary black hole has a smaller effect on the waveform, and accordingly we find that $\cos\theta_2$ is less informative than any of the other tested parameters, with $\SDR[\cos\theta_2]$ clustered near 0 (not shown).

Gravitational-wave detectors are sensitive in a given frequency range, and which part of the waveform it corresponds to (inspiral, merger) depends on the mass of the binary. This suggests a follow-up question: is spin misalignment equally observable for binaries of different masses?
$\chip$ and $\beta$ were chosen explicitly as precession parameters, controlling the average rate of precession and the size of the precession-induced amplitude modulations, respectively.
We may therefore expect that they are more informative for low-mass systems, which spend more cycles in the detector band and have a clearer separation of orbital and precession timescales.

\begin{figure}
    \centering
    \includegraphics[width=\linewidth]{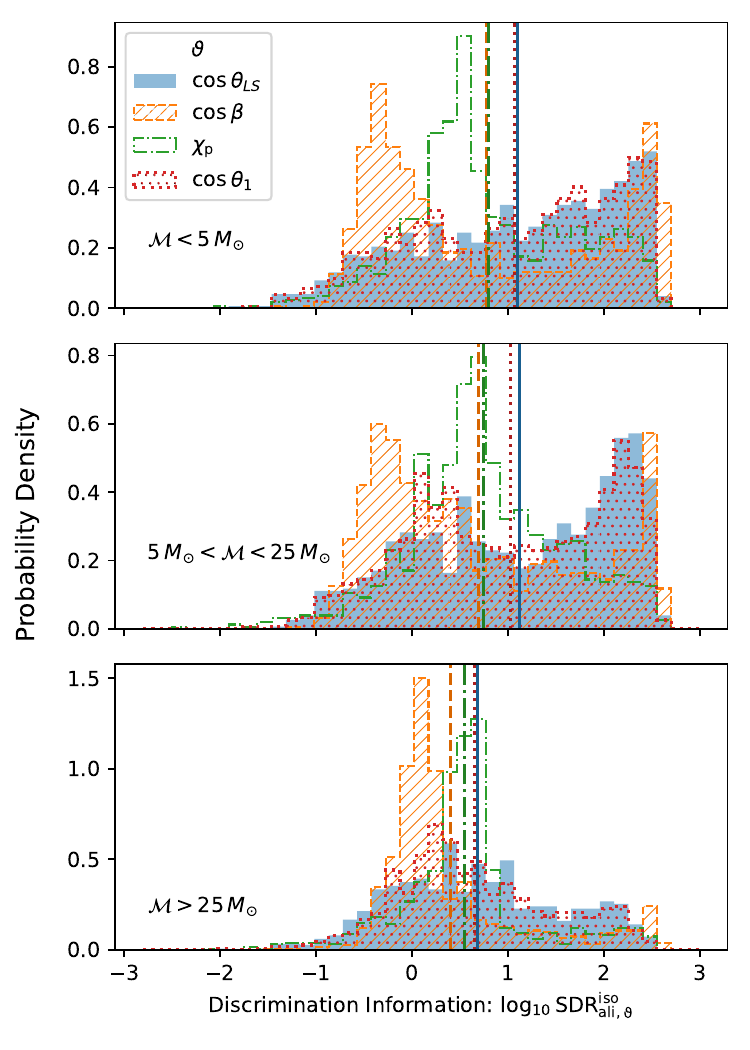}
    \caption{
    Distribution of \SDR for the four parameters $\vartheta~\in~\{\ctls, \cos\beta, \chip, \cos\theta_1\}$ in synthetic events across three chirp-mass ranges.
    These are the same histograms as in the right panel of Fig.~\ref{fig:savage-dickey}, except disaggregated by detector-frame chirp-mass.
    Savage--Dickey ratios based on $\ctls$ appear to be less sensitive to increases in mass than those based on $\cos\beta$ or $\chip$.
    Notably, the distributions of Savage--Dickey ratios for $\cos{\theta_{LS}}$ as well as $\cos{\theta_1}$ remain relatively flat  across the three mass ranges. Additionally, the average $\log\SDR[\cos{\theta_{LS}}]$ value is consistently higher than the average $\log\SDR$ of the other parameters.}
    \label{fig:injection_sdrs_mass_disbs}
\end{figure}

We explore this in Fig.~\ref{fig:injection_sdrs_mass_disbs}, which disaggregates the simulated \SDR into three chirp-mass bins (1--\SI{5}{M_\odot}, 5--\SI{25}{M_\odot}, 25--\SI{125}{M_\odot}).
Indeed, we find that low-mass sources are more informative than high-mass ones for all four choices of $\vartheta$: low-mass systems have a distribution of $\log\SDR$ that is broader and with a larger mean.
The mean discrimination information of $\cos{\theta_{LS}}$ is consistently higher than that of $\cos{\beta}$ and $\chi_{\rm p}$ across the three chirp-mass groups, as also reported in Table~\ref{table:DKL_values}.
For the low and high chirp mass bins, $\tilde{D}_{{\rm KL},\ctls}$ and $\tilde{D}_{{\rm KL},\cos\theta_1}$ are approximately equal within the sampling error.
For the mid chirp mass range, $\ctls$ is $\sim9\%$ more informative than $\cos{\theta_1}$.
To interpret these numbers, recall that $\tilde{D}_{{\rm KL},\vartheta}$ is inversely proportional to the average number of detections required to rule out the aligned-spin hypothesis at a given significance if the true population follows the isotropic-spin model, if one only considered the information in the $\vartheta$ posteriors.

\begin{table}[t]
\caption{ $\tilde{D}_\mathrm{KL,\vartheta}\infdivx{p(d | \Hiso)}{p(d | \Hali)}$, the mean discrimination information between the isotropic- and the aligned-spin hypotheses, using each parameter
$\vartheta \in \{
    \ctls, \allowdisplaybreaks
    \cos\beta, \allowdisplaybreaks
    \chip, \allowdisplaybreaks
    \cos\theta_1
\}$, see Eq.~\eqref{eq:dkl_1par}.
Error bars report the standard deviation, estimated using the bootstrap method by resampling Eq.~\eqref{eq:dkl_1par_approx}.}

\label{table:DKL_values}
\begin{ruledtabular}
\begin{tabular}{c|cccc}
$\mathcal M$ range & $\tilde{D}_{\mathrm{KL},\ctls}$ & $\tilde{D}_\mathrm{KL,\cos\beta}$ & $\tilde{D}_\mathrm{KL,\chip}$ & $\tilde{D}_\mathrm{KL, \cos\theta_1}$ \\

(M$_\odot$) & (bits) & (bits) & (bits) & (bits) \\[2pt]
\hline\rule{0pt}{10pt}
1--5 & $3.65\pm0.11$ & $2.65\pm0.09$ & $2.58\pm0.12$ & $3.53\pm0.11$ \\
5--25 & $3.72\pm0.11$ & $2.49\pm0.08$ & $2.31\pm0.11$ & $3.39\pm0.11$ \\
25--125 & $2.27\pm0.09$ & $1.81\pm0.07$ & $1.34\pm0.08$ & $2.17\pm0.08$ \\[2pt]
All & $3.21\pm0.06$ & $2.32\pm0.05$ & $2.08\pm0.06$ & $3.03\pm0.06$ \\
\end{tabular}
\end{ruledtabular}
\end{table}

\begin{figure}
    \centering
    \includegraphics[width=\linewidth]{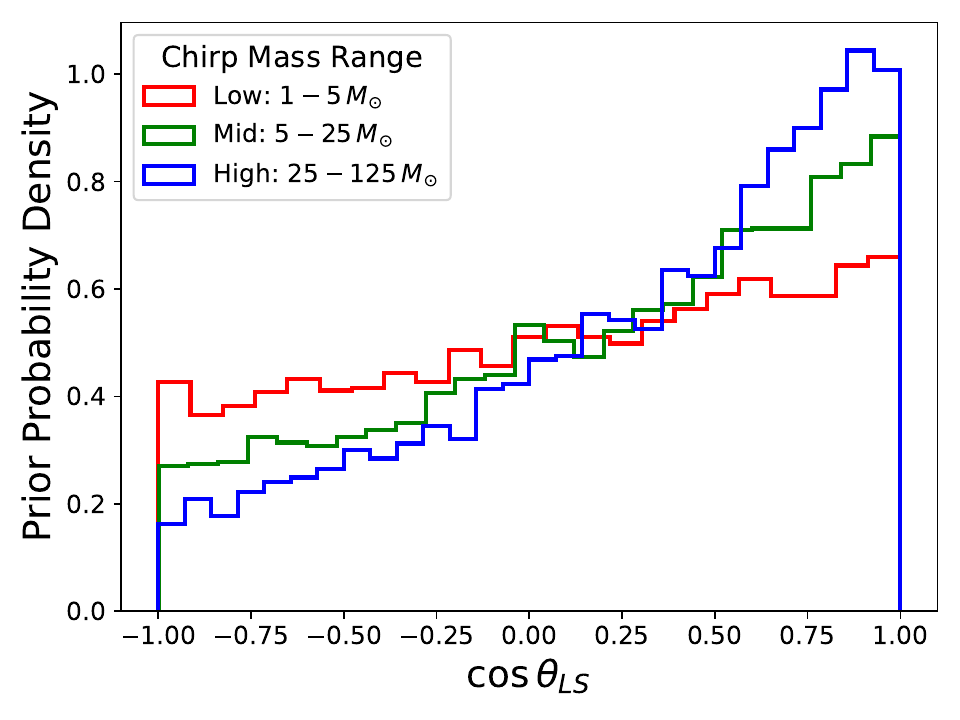}
    \caption{Mass-dependent selection effects on spin--orbit tilt $\theta_{LS}$. The data were initially generated according to a volumetric spin prior, i.e.\ isotropic (uniform in $\ctls$) and favoring high spin magnitudes $p(\chi) \sim \chi^2$, and then filtered to select events detectable by LVK as described in Section~\ref{sec:performing_parameter_test}.
    As the mass increases, the bias towards observing aligned $\bm{S}$ and $\bm{L}$ increases.
    We expect spin selection effects to be less pronounced for priors that favor smaller spin magnitudes.
    }
    \label{fig:selection_effs}
\end{figure}

Another consequence of the detectors being sensitive to different portions of the waveform are mass-dependent spin-selection effects.
As described above, the isotropic spin prior used would result in a flat prior in $\cos{\theta_{LS}}$, except that, when accounting for the detectability criterion, it becomes biased towards aligned $\cos{\theta_{LS}}$ at higher masses.
This is due to the so-called orbital hangup effect \citep{Campanelli2006}, which delays the merger for systems with a high spin aligned with the orbit, causing stronger gravitational wave emission and enhancing detectability.
We show this effect in Fig.~\ref{fig:selection_effs}, where we observe that selection effects indeed favor aligned-spin configurations and the effect is more pronounced for more massive systems, for which the detectors are most sensitive to the merger.

\section{Spin misalignment in GWTC-5.0 events}
\label{sec:LVK}

The values of the mean discrimination information $\tilde{D}_{\rm KL,\vartheta}$ in Table~\ref{table:DKL_values} support that, of the tested parameters, $\ctls$ is the most informative about spin--orbit alignment.
In this section, we reanalyze the posterior distributions of individual events from GWTC-5.0 in terms of $\ctls$.

We begin by computing the Savage--Dickey ratio for $\vartheta \in \{ \ctls, \chip, \cos\beta, \cos\theta_1\}$ of each event using the GWTC-5.0 posterior distributions \citep{lvk_gwtc2p1pe_zenodo,lvk_gwtc3pe_zenodo,lvk_gwtc4p1_zenodo,LVK_Zenodo,lvk_gwtc5pe_zenodo_part1,lvk_gwtc5pe_zenodo_part2} and a set of draws from the prior distribution under the isotropic spin model. For events detected in O4a and O4b, we use samples obtained using the \texttt{NRSur7dq4} \citep{Varma2019} approximant when available and \texttt{IMRPhenomXPHM-SpinTaylor} \citep{Pratten2021, Colleoni2025} otherwise. For events detected in O1--O3b, we use samples obtained using the \texttt{IMRPhenomXPHM} \citep{Pratten2021} approximant.
Unless stated otherwise, in this section we adopt a prior uniform in spin magnitudes instead of the volumetric spin prior used in Section~\ref{sec:performing_parameter_test}, to match the prior used by LVK to generate the posteriors, and specify the spins at a reference frequency of \SI{20}{\hertz}.

\begin{figure}
    \centering
    \includegraphics[width=\linewidth]{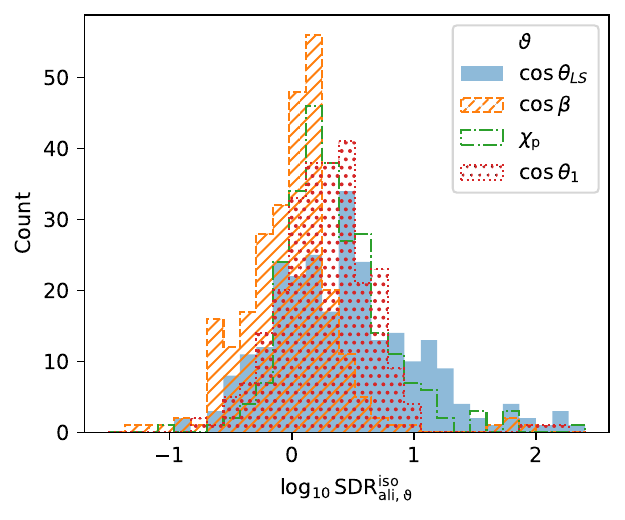}
    \caption{Distribution of $\SDR$ values for the 259 binary black holes in GWTC-5.0. Notably, the distribution of $\SDR[\cos\theta_1]$ is not as well populated as that of $\SDR[\cos\theta_{LS}]$ at the high end of the tail. Furthermore, $\SDR[\cos\beta]$ extends to more negative values than any other parameter.}
    \label{fig:lvk_bayes_factors}
\end{figure}

Figure~\ref{fig:lvk_bayes_factors} shows the Savage--Dickey ratios between the isotropic- and aligned-spins hypotheses for events in GWTC-5.0.
We observe that $\log\SDR[\ctls]$ has broader tails than $\log\SDR[\chip]$ or $\log\SDR[\cos\theta_1]$ at both the positive and negative ends, confirming that it generally contains more information supporting or rejecting the aligned-spin hypothesis for different events.
$\SDR[\cos\beta]$ features a steeper tail at the positive end and broader at the negative.

\begin{figure}
    \centering
    \includegraphics[width=\linewidth]{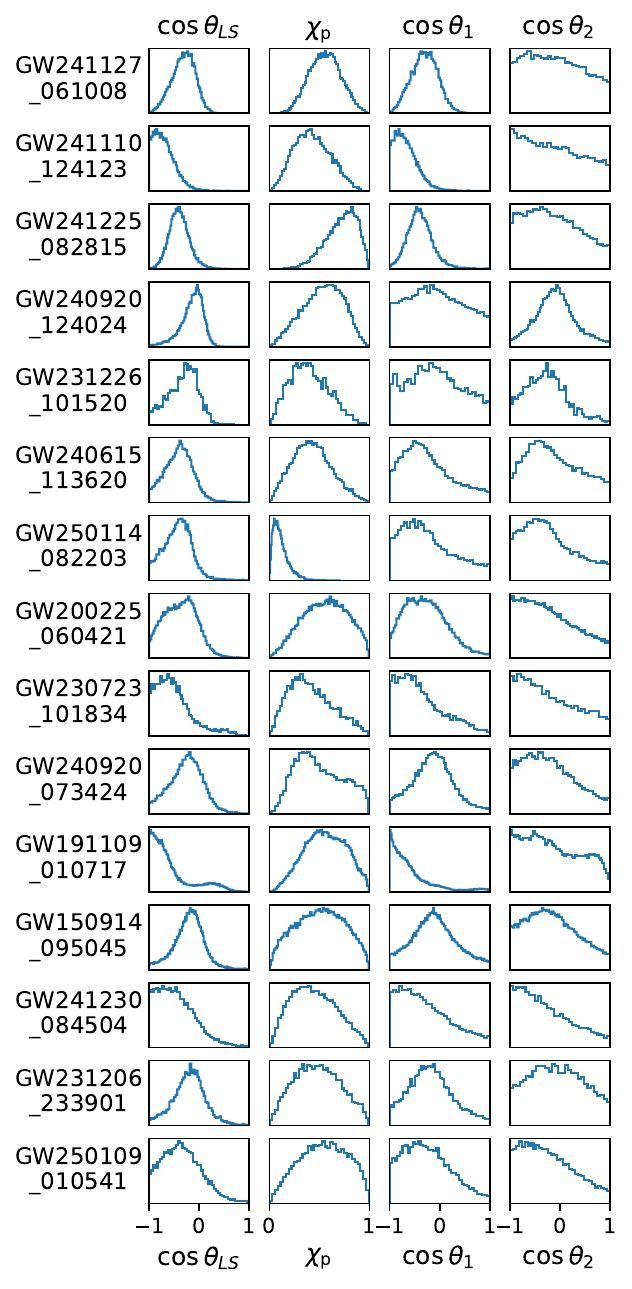}
    \caption{The fifteen GWTC-5.0 events with highest $\SDR[\ctls]$ values, ruling out the aligned total-spin configuration $\ctls=1$ to varying degrees. The total-spin tilt is usually better constrained than either component-spin tilt.
    }
    \label{fig:10_lowest_sdrs}
\end{figure}

Next, we sort the events by decreasing \SDR[\ctls], as a systematic way to identify the events with most evidence of spin--orbit misalignment:
The events with the highest $\SDR[\ctls]$ values are those most likely to have spin--orbit misalignment, while those with the lowest $\SDR[\ctls]$ are the ones most likely to have their spin aligned with the orbit.
We present a more in-depth discussion of events with spin misalignment by focusing on the 15 events with the highest \SDR[\ctls] in Fig.~\ref{fig:10_lowest_sdrs}.
The posterior distributions of $\cos{\theta_{LS}}$, $\chi_{\rm p}$, and the individual spin tilts $\cos{\theta_1}$ and $\cos{\theta_2}$ are shown.
Notably, the $\cos{\theta_{LS}}$ posterior distributions of these events strongly disfavor spin--orbit alignment, having little to no support at $\cos{\theta_{LS}}=1$, and are generally better constrained than the individual spin tilts $\cos{\theta_1}$ and $\cos{\theta_2}$ (save that in events where the component masses are unequal, the primary spin usually dominates the total spin and $\theta_1 \approx \theta_{LS}$, see Eq.~\eqref{eq:costhetals}).
In contrast, in most cases it is difficult to place strong constraints on the in-plane spin components through $\chi_{\rm p}$ due to the broad posterior distributions.
Conversely, the events with low \SDR[\ctls] have posterior distributions peaked at $\ctls=1$ \footnote{The posterior distributions of each event can be found at \url{https://github.com/charlie-gibson/cos_theta_ls}.}.

To discuss these events, we classify them into three categories: (i) events for which precession can be ascertained via commonly used parameters even without resorting to $\ctls$, (ii) those where $\ctls$ rules out spin--orbit alignment however they are consistent with having no spin, and (iii) those where precession can be identified more easily through $\ctls$. 

In the first category, we find that the top three events in terms of total-spin misalignment (GW241127\_061008, GW241110\_124123, and GW241225\_082815; Fig.~\ref{fig:10_lowest_sdrs}) are also identifiable as precessing through their primary spins, which are confidently tilted ($\cos\theta_1 \neq 1$).
In fact, GW241127\_061008 and GW241110\_124123 have unequal masses and thus have $\theta_1 \approx \theta_{LS}$.
GW241127\_061008 and GW241225\_082815 also have $\chip \neq 0$.

\begin{figure}
    \centering

    \includegraphics[width=\linewidth]{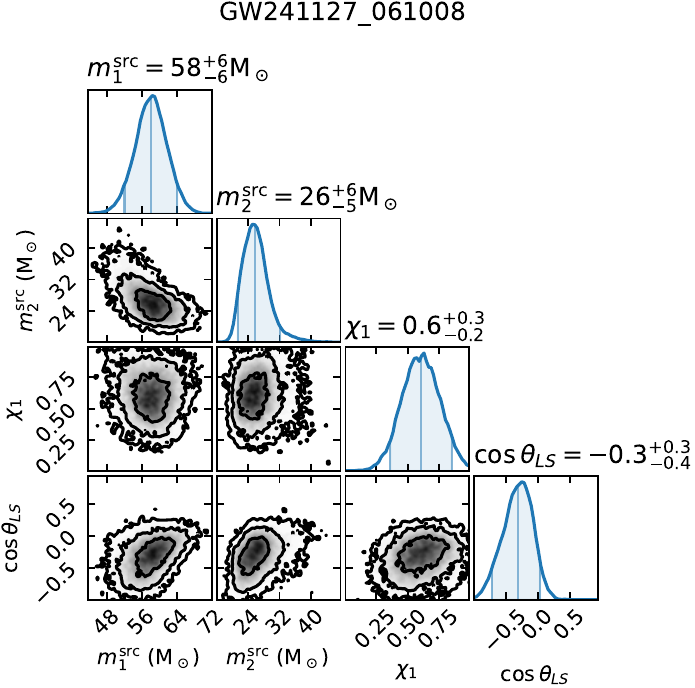}
    \caption{Masses, primary spin magnitude, and total-spin tilt of GW241127\_061008. This event, having the greatest value of $\SDR[\ctls]=191.3$, has properties that point to it being a second-generation black hole merger.
    Its primary mass may fall within the pair-instability supernova mass gap, and it has a large primary spin consistent with $\chi_1 \approx 0.7$ and significantly tilted with the respect to the orbital angular momentum ($\theta_1 > 73^\circ$ at 99\% credibility). Although not shown, $\chi_{\rm p} > 0.22$ at 99\% credibility. $\chi_2$ is unconstrained, and $\chi_{\rm eff}$ is consistent with 0. Contours enclose 50\%, 90\% or 99\% of the probability.}
    \label{fig:gw241127_cornerplot}
\end{figure}

Remarkably, the event with highest $\SDR[\ctls]$, GW241127\_061008, is a very compelling case for a second-generation merger, to our knowledge not previously discussed as such.
As shown in Fig.~\ref{fig:gw241127_cornerplot}, the primary spin is constrained away from 0 and consistent with $\chi_1 \approx0.7$ (as expected for a merger remnant \cite{Pretorius2005, Scheel2009}), and the mass ratio is constrained away from unity and consistent with $m_2/m_1 \approx 0.5$ (a merger remnant would typically have roughly twice the mass of first-generation black holes merging in its host cluster \cite{Gerosa2021b}).
Moreover, it is difficult to explain the properties of GW241127\_061008 within an isolated binary evolution scenario, because its primary spin is significantly tilted with respect to the orbit ($\theta_1 > 73^\circ$ at 99\% credibility).
Finally, its primary mass ($m_1^\mathrm{src} = 58 \pm 10 \, \mathrm{M}_\odot$ at 99\% credibility) may be in the pair-instability supernova mass gap, expected in the range $\sim 60$--$\SI{135}{M_\odot}$ in the mass function of black holes formed from stellar collapse but possibly populated by merger remnants \cite{Fowler1964, Farmer2020, Mehta2022, Farag2022}.
The location of the lower edge of the mass gap is currently predicted at $60^{+32}_{
-14}\,\mathrm{M}_\odot$ (at $3\sigma$), with the main source of uncertainty being the $^{12}\mathrm{C}\,(\alpha, \gamma)\,{}^{16}\mathrm{O}$ nuclear reaction rate \cite{Farag2022}.
The theoretical uncertainty is wide enough that GW241127\_061008's posterior on the primary mass could lie entirely in the pair-instability supernova gap, thus, an improved understanding of the $^{12}\mathrm{C}\,(\alpha, \gamma)\,{}^{16}\mathrm{O}$ rate could exclude a stellar-collapse origin for this black hole.
Observationally, gravitational wave detections suggest a consistent but generally lower edge of the gap at $44^{+5}_{-4}\,\rm M_\odot$ \citep{Tong2026}, which would favor the interpretation that GW241127\_061008's primary is in the mass gap.
All in all, GW241127\_061008 is a strong case for a second-generation merger, which would imply a formation in a dense environment such as a nuclear or globular cluster.
Other second-generation candidates discussed in the literature are GW241110 (second in terms of $\SDR[\ctls]$, see Fig.~\ref{fig:10_lowest_sdrs}) and GW241011, which are also consistent with a tilted primary spin of $\approx 0.7$ and unequal masses \cite{Abac2025b}; both have a lower mass $m_1 \approx \SI{20}{M_\odot}$.
A hierarchical merger has also been proposed to explain the high masses and spins of GW231123 \cite{Abac2025, Passenger2026}. Because its masses and spins are generally larger than expected from first-generation merger remnants \cite{Croon2026}, that hypothesis requires invoking even higher-generation mergers, or original progenitors with high, aligned spin \cite{Stegmann2025}.

In the second category, we identify the events GW240920\-\_124024, GW231226\-\_101520, GW240615\-\_113620, GW250114\-\_082203, GW200225\-\_060421, GW230723\-\_101834, GW150914\-\_095045, GW241230\-\_084505, GW231206\-\_233901, and GW250109\-\_010541.
These systems have relatively well localized posteriors that exclude $\ctls=1$ and therefore exclude spin--orbit alignment.
However, as we expand below, these events also allow both black holes to be nonspinning, meaning that we cannot make claims of precession or rule out most formation scenarios for them.

\begin{figure}
    \centering
    \includegraphics[width=\linewidth]{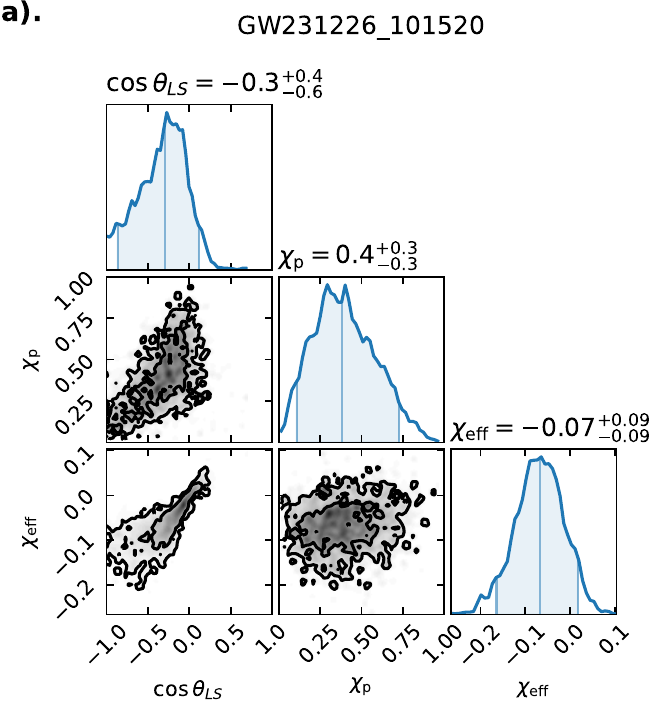}

    \vspace{3pt}
    \includegraphics[width=0.59\linewidth]{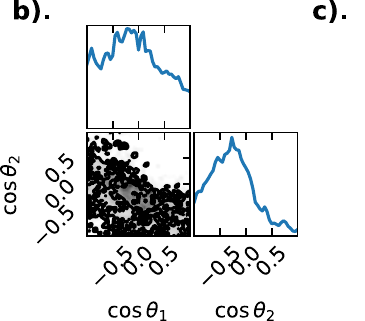}
    \hspace{-10pt}\includegraphics[width=0.43\linewidth]{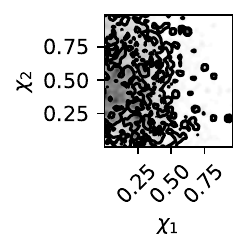}
    \caption{Corner plots of selected spin parameters for GW231226\_101520, which had the highest $\SDR[\ctls]$ of all GWTC-4.0 events despite having unremarkable posteriors for the effective spins ($\chip$, $\chieff$), individual spin tilts ($\theta_1$, $\theta_2$), and spin magnitudes ($\chi_1$, $\chi_2$).
    \textbf{a).} From the 1D posteriors, the aligned configuration $\ctls=1$ is rejected, even though $\chip=0$ or $\chieff>0$ are allowed.
    Looking at the 2D correlations, we see that the $\chip=0$ solution corresponds to the anti-aligned configuration ($\ctls=-1$), and $\chieff=0$ to a spin perpendicular to the orbit ($\ctls\approx0$).
    \textbf{b).} The component spins are individually consistent with being aligned ($\cos\theta_{1, 2}=1$), but, notably, not simultaneously.
    \textbf{c).} However, we cannot exclude the possibility that both spins are very small, which is surprising given the other constraints.
    }
    \label{fig:gw231226_cornerplot}
\end{figure}

In Fig.~\ref{fig:gw231226_cornerplot} we provide a more in-depth analysis on GW231226\_101520, which we identify as the event with highest \SDR[\ctls] in GWTC-4.0.
It was highlighted in that catalog as one of the loudest signals, and one of the events with most support for $\chieff < 0$ (93\% under the fiducial parameter estimation prior) \citep{GWTC4}.
Interestingly, despite completely rejecting alignment of the total spin with the orbit , this event has rather unremarkable marginal posteriors for many of the main spin-related parameters: the precessing effective spin $\chip$ is unconstrained, the aligned effective spin $\chieff$ could be either positive, zero, or negative, the individual spin tilts $\theta_1$ and $\theta_2$ could range anywhere from complete alignment to complete anti-alignment, and the spin magnitudes have no constraints other than disfavoring a high primary spin ($\chi_1 < 0.63$ at 95\% credibility).
However, when we consider the \textit{joint} posterior on $\chip$ and $\ctls$ (Fig.~\ref{fig:gw231226_cornerplot}a), we see that the $\chi_{\rm p}=0$ limit does not correspond to having aligned spin ($\cos{\theta_{LS}}=1$), but anti-aligned spin ($\cos{\theta_{LS}}=-1$). 
Looking at $\chi_{\rm eff}$ and $\chi_{\rm p}$, it becomes clear that a positive $\chieff$ is only allowed if $\chip$ is also away from 0, that is, at least one spin is tilted.
The individual spins of GW231226\_101520 are individually allowed to be aligned, but not simultaneously (Fig.~\ref{fig:gw231226_cornerplot}b).
An immediate interpretation would be that the system must have some level of misalignment; however, both black holes could have negligible spins (Fig.~\ref{fig:gw231226_cornerplot}c).

\begin{figure}
    \centering
    \includegraphics[width=.9\linewidth]{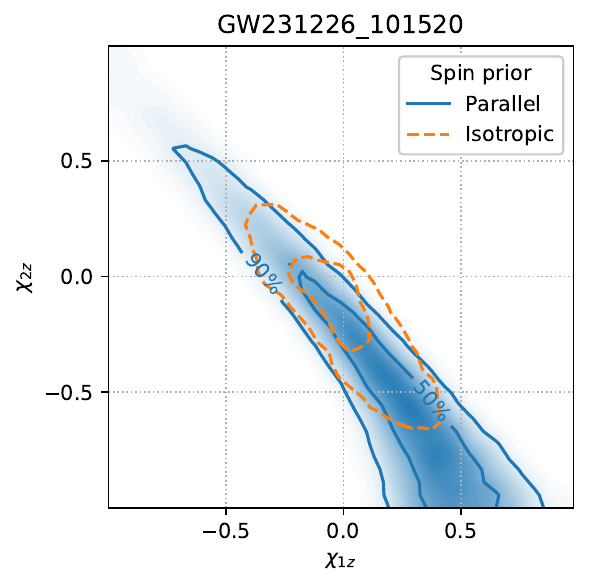}
    \caption{Reconciling the observations that GW231226\_101520 is consistent with zero spins, but not aligned spins.
    Assuming no in-plane spins, the posterior only just accommodates zero spins, but rejects aligned spins ($\chi_{1z}, \chi_{2z}>0$) that are not negligibly small.
    }
    \label{fig:GW231226_101520_parallel_spins}
\end{figure}

The conclusions that GW231226\_101520 can be nonspinning but cannot have aligned spins are somewhat puzzling, since no spins is the limiting case of small aligned spins.
To investigate this,  we reanalyze the event under three different spin priors: isotropic tilts and uniform spin magnitudes; zero spins; or spins parallel (meaning aligned or anti-aligned) to the orbit with a prior uniform in $\chieff$.
We use the inference software \texttt{cogwheel} \cite{Roulet2022, Roulet+24} and the waveform model \texttt{IMRPhenomXODE} \cite{Yu2023}.
We confirm the findings above: the isotropic-spins model describes GW231226\_101520 just as well as the nonspinning model (Bayes factor $\mathcal B^\iso_\text{no spin} \approx 1.25$), marginally better than the parallel-spins prior ($\mathcal B^\iso_\parallel \approx 3.25$), but significantly better than restricted to aligned spins ($\mathcal B^\iso_\ali \approx \num{1.98e3}$).
This is because 99.96\% of the parallel-spins posterior supports at least one component spin anti-aligned with the orbit (Fig.~\ref{fig:GW231226_101520_parallel_spins}).

\begin{figure}
    \centering
    \includegraphics[width=\linewidth]{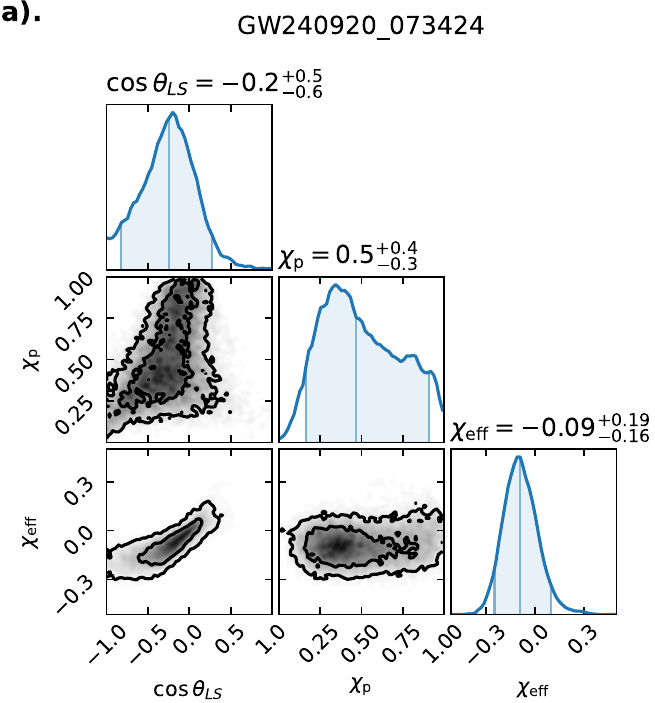}

    \vspace{3pt}
    \includegraphics[width=0.55\linewidth]{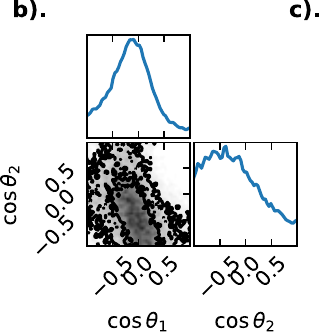}%
    \includegraphics[width=0.45\linewidth]{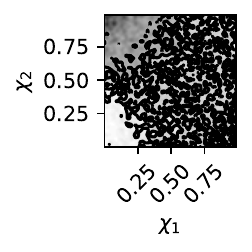}
    \caption{Corner plots of selected spin parameters for GW240920\_073424. This event is categorized as one that has spin--orbit misalignment that may not easily be seen by existing misalignment parameters.
    \textbf{a).} From the 1D posteriors, the aligned configuration $\ctls=1$ is rejected, even though $\chip=0$ or $\chieff>0$ are allowed.
    Looking at the 2D correlations, we see that the $\chip=0$ solution corresponds to the anti-aligned configuration ($\ctls=-1$).
    \textbf{b).} The component spins are individually consistent with being aligned ($\cos\theta_{1, 2}=1$), but, notably, not simultaneously.
    \textbf{c).}
    A scenario where both spins are small simultaneously is disfavored by the data, unlike the example in Fig.~\ref{fig:gw231226_cornerplot}.
    }
    \label{fig:gw240920_cornerplot}
\end{figure}

Finally, we identify the events GW240920\_073424 and GW191109\_010717 whose $\ctls$ posteriors show evidence of precession that is not apparent using existing parameters.
The posteriors for both of these events favor at least one black hole to be spinning.
Figure~\ref{fig:gw240920_cornerplot} shows posterior distributions for the relevant spin parameters of GW240920\_073424.
Similarly to GW231226\_101520 (see Fig.~\ref{fig:gw231226_cornerplot}), GW240920\_073424 excludes $\ctls=1$ but has unremarkable marginal posteriors for $\chip$, $\chieff$ and the spin tilts: $\chip, \theta_1$ and $\theta_2$ are poorly constrained, and $\chieff$ is consistent with 0.
These facts make it unlike the events in the first category which showed clearer signs of spin--orbit misalignment via the standard parameters.
Here too, the $\chip=0$ configuration only corresponds to anti-aligned total spin, $\ctls=-1$ (Fig.~\ref{fig:gw240920_cornerplot}a).
Finally, we see that at least one black hole has to be spinning (Fig.~\ref{fig:gw240920_cornerplot}c), distinguishing GW240920\_073424 also from the aforementioned events with $\ctls$ distributions that rule out spin--orbit alignment but could be nonspinning.

The case of GW191109\_010717 is qualitatively similar, in that neither $\chieff$, $\chip$, $\theta_1$ or $\theta_2$ posteriors alone rule out aligned spins but $\theta_{LS}$ does, and the spins cannot be simultaneously zero (although the distribution of $\cos\theta_1$ is roughly similar to that of $\ctls$, the distribution of $\ctls$ more strongly rejects spin--orbit misalignment, see Fig.~\ref{fig:10_lowest_sdrs}; also note that the event has been highlighted as being among those with largest support for $\chieff<0$, 90\% under the fiducial isotropic prior \citep{GWTC3}).
However, it should be noted that GW191109\_010717 has data quality issues that may have affected the inference results, in particular about spin \citep{Udall+25, Kumar+26}, and that waveform model systematics also are of concern \citep{GWTC3, Islam2025}.

\section{Astrophysical distribution of total-spin tilts}\label{sec:population}

In Section~\ref{sec:LVK} we have highlighted several events with signs of spin misalignment, explicitly selected out of a large catalog for this reason.
In this section, we characterize the distribution of total spin tilts of the full GWTC-5.0 catalog through hierarchical Bayesian inference \cite[e.g.,][]{Mandel2019} on a set of simple parametric models.

\subsection{Data and models}

We analyze 259 gravitational wave events from binary black hole systems with a $\text{false-alarm rate} \leq \SI{1}{yr^{-1}}$ in the GWTC-5.0 catalog \citep{GWTC5} (excluding GW190814).
We use the publicly released posterior samples on Zenodo \citep{lvk_gwtc2p1pe_zenodo, lvk_gwtc3pe_zenodo, lvk_gwtc4p1_zenodo, lvk_gwtc5pe_zenodo_part1, lvk_gwtc5pe_zenodo_part2} and the same choice of waveform approximants as described in Section \ref{sec:LVK}. We use the injections available at \cite{lvk_gwtc5_injections} (details of which can be found at \cite{Essick2025}) to account for selection effects in our analysis. 

Given our observation in Sec.~\ref{sec:LVK} that events consistent with zero spin can exclude $\ctls=1$, we include a nonspinning subpopulation in the model with a mixture fraction of $f_\text{no-spin}$, so that those events can be accounted for without invoking misaligned spins.
To avoid pathological importance-sampling weights, the spin magnitudes for the nonspinning subpopulation are modeled as truncated Gaussians with a mean fixed at $\mu_\text{no-spin}=0$ and a fixed width of $\sigma_\text{no-spin}=0.05$, with isotropically distributed orientations.
For the spinning subpopulation, we assume the spin magnitudes are independently and identically distributed draws from a truncated Gaussian distribution whose mean and width are free parameters.
We model the $\ctls$ distribution of the spinning subpopulation as a mixture of an isotropic component (flat in $\ctls$) and a truncated Gaussian distribution, 
\begin{multline}
    p_{\rm pop}(\ctls, \chi_1, \chi_2 \mid f_\text{no-spin}, f_{\rm iso}, \mu_{LS}, \sigma_{LS}, \mu_\text{spin}, \sigma_\text{spin})\\
    =\frac{f_\text{no-spin}}{2} \times p_\text{no-spin}(\chi_1, \chi_2 \mid \mu_\text{no-spin}, \sigma_\text{no-spin})\\
    + (1-f_\text{no-spin})\times p_\text{spin}(\chi_1, \chi_2 \mid \mu_\text{spin}, \sigma_\text{spin})\times\\
    \left[ \frac{f_{\rm iso}}{2}
    + (1-f_{\rm iso}) \mathcal{N}_{[-1, 1]}(
        \ctls \mid \mu_{LS}, \sigma_{LS})\right], \label{eq:population_model_costhetals}
\end{multline}
where $p_\text{no-spin}, p_\text{spin}$ are the corresponding spin magnitude distributions as described above and $f_\text{no-spin}, f_{\rm iso}$, $\mu_{LS}$, $\sigma_{LS}, \mu_\text{spin},  \sigma_\text{spin}$ are the hyperparameters whose posteriors we aim to infer.

We consider three nested models corresponding to progressively relaxed assumptions about $\ctls$ distribution of the model in Eq.~\eqref{eq:population_model_costhetals}.
First we consider the \texttt{preferentially aligned} model, in which the entire population has spin tilts preferentially aligned with the orbital angular momentum, by setting $f_{\rm iso}=0$ and $\mu_{LS}=1$, while $\sigma_{LS}$ is left as a free parameter that controls the degree of misalignment. Second, we consider a \texttt{preferentially aligned + isotropic} model where we allow $f_{\rm iso}$ to be a free parameter (while setting $\mu_{LS}=1$). Finally, we consider the \texttt{Gaussian + isotropic} model allowing all three parameters $f_{\rm iso}, \mu_{LS}, \sigma_{LS}$ to be free. For the mass distribution, we adopt the \texttt{Broken Power Law + 2 Peaks} model \citep{PopulationO4a}. We model the redshift distribution with a single power law $\propto(1+z)^\kappa$. We jointly infer the distribution of the mass, spins, and redshift, and sample the posterior using nested sampling with \textsc{dynesty} \citep{dynesty}.

\subsection{Results}

\begin{figure}
    \centering
    \includegraphics[width=\linewidth]{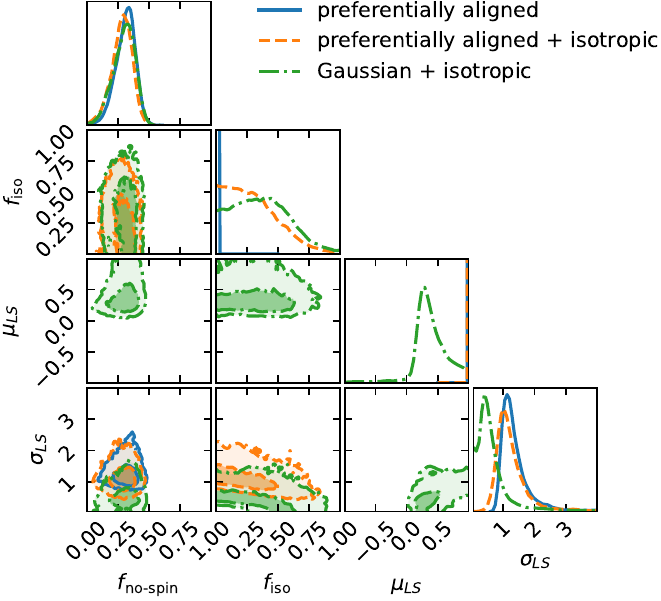}
    \caption{The posterior distribution of the parameters describing the spin tilt distribution 
    in the three nested models: \texttt{preferentially aligned}, \texttt{preferentially aligned + isotropic}, \texttt{Gaussian + isotropic} obtained by performing hierarchical Bayesian inference on the binary black holes in GWTC-5.0.}
    \label{fig:population_cornerplot}
\end{figure}

Figure \ref{fig:population_cornerplot} shows the posterior distribution for the hyperparameters of the three $\ctls$ models described above along with the fraction of nonspinning systems $f_\text{no-spin}$. Across all three models, we find that the data favor a broad distribution in $\ctls$. Specifically, in the \texttt{preferentially aligned} model, we recover a width of $\sigma_{LS} = 1.2 ^{+1.1}_{-0.3}$ in the $\cos{\theta_{LS}}$ distribution, indicating significant spin--orbit misalignment in the population. Allowing for an isotropic component in the \texttt{preferentially aligned + isotropic} model, we find that although narrower distributions in $\ctls$ are allowed in the preferentially-aligned component, models with small $\sigma_{LS}$ require a large fraction of the population to have isotropic spins. In the \texttt{Gaussian + isotropic} case, we find that the mean of the Gaussian peaks away from 1, $\mu_{LS} = 0.4 ^{+0.5}_{-0.3}$. All three models find a similar fraction of non-spinning systems, $f_\text{no-spin} = 0.3^{+0.1}_{-0.2}$. 

These results are in agreement with findings from other works \citep{VitaleBiscoveanuTalbot22, Callister2024, Stegmann2026, PopulationO4a}, that explored similar models but for the individual spin tilts $\theta_1, \theta_2$ rather than the total spin tilt $\theta_{LS}$. Those works also found a peak at a small $\cos\theta \approx 0.2$, which \citet{Stegmann2026} interpret as evidence for mergers induced through hierarchical triple evolution, that predicts a preference for spin tilts perpendicular to the orbital angular momentum. However, we find that introducing these additional free parameters does not significantly improve the fit to the data. The Bayes factors between the \texttt{preferentially aligned} and \texttt{Gaussian + isotropic} models are close to unity. These models are favored over the \texttt{preferentially aligned + isotropic} model which has a Bayes factor of  approximately 0.5 with respect to the other two.
Therefore, we find that the evidence for formation in triples remains inconclusive with GWTC-5.0, as \citet{Wolfe2026} concluded from GWTC-4.0.
Alternatively, the peak at near in-plane spins could be spurious and more events would be needed to better constrain the true distribution \cite{VitaleMould2025}. 

Next, we investigate whether the observed population is consistent with binaries formed through isolated binary evolution, in which spin--orbit misalignment would arise solely from the supernova kick imparted during the formation of the secondary black hole. Following \citet{Callister2021}, we simulate $5 \times 10^{4}$ binaries with kick velocities drawn from a Maxwellian distribution and imparted isotropically to the secondary black hole (refer to Section~3.1 of \cite{Callister2021} for details). As in \citet{Callister2021}, we assume that the first born black hole is the primary (more massive one) and has no spin. The secondary black hole has some spin aligned with the initial orbit and accrued via tidal torquing of its progenitor helium core. 
We consider two prescriptions for the natal kick imparted to the secondary black hole, a low kick velocity distribution, with $\sigma_{\rm kick}^{\rm low} = \SI{250}{\kilo\meter/\second}$ and a high kick velocity distribution $\sigma_{\rm kick}^{\rm high} = \SI{1000}{\kilo\meter/\second}$, and compute the resulting $\ctls$ distribution of the merging binary black hole systems.

\begin{figure}
    \centering
    \includegraphics[width=\linewidth]{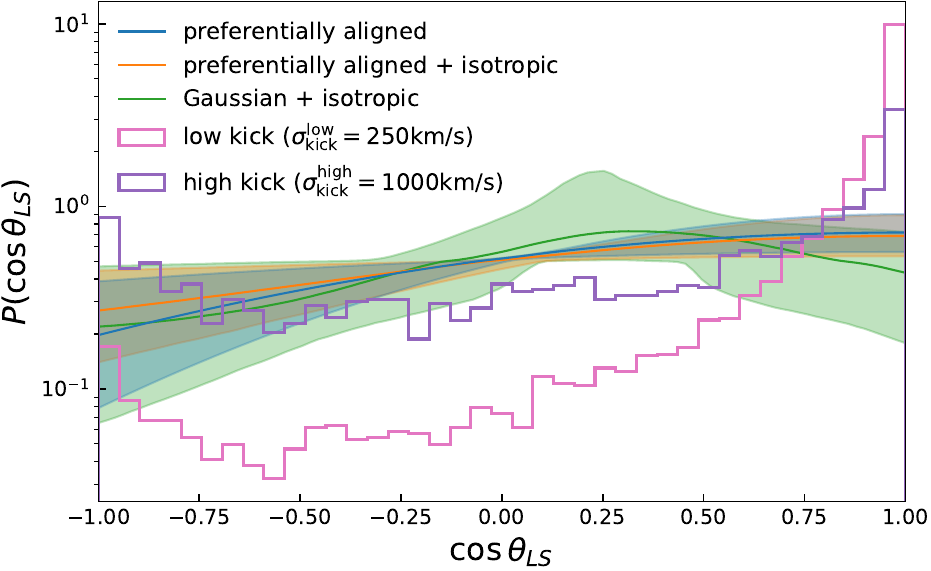}
    \caption{Posterior predictive distributions of the \texttt{preferentially aligned} (blue), \texttt{preferentially aligned + isotropic} (orange) and \texttt{Gaussian + isotropic} (green) models. The histograms are the simulated binary black hole systems formed via the isolated formation channel with kick velocities drawn from a Maxwellian distribution. We show two kick-velocity prescriptions, favoring low (pink) or high (purple) kick velocities.}
\label{fig:costhetals_predictive_distribution}
\end{figure}

In Fig.~\ref{fig:costhetals_predictive_distribution}, we compare these simulated distributions to the population distributions inferred from the models introduced above. The simulated populations predict substantially more aligned systems than is favored by the data. Increasing the magnitude of kick velocities reduces this discrepancy somewhat, leading to a broader tilt distribution.
However, we note that our parametrized models are not sufficiently flexible to fully reproduce the simulated $\ctls$ distributions. In particular, our most general model (\texttt{Gaussian + isotropic} model) assumes a single Gaussian component for the aligned population and therefore cannot capture the secondary peak near $\ctls=-1$, which arises from kicks imparted nearly opposite to the orbital velocity. Although our models fail to describe this formation mechanism, the \texttt{preferentially aligned} and \texttt{preferentially aligned + isotropic} may still possibly provide the best fit to the high-kick scenario. The low-kick scenario remains ruled out, as in \citet{Callister2021}, although it could still account for a fraction of the population.

\section{Conclusion}\label{sec:discussion}

In this work we accomplished three main goals.
Through a statistical framework based on hypothesis discrimination, we determined that \textbf{the angle $\theta_{LS}$ between the total spin and the orbital angular momentum is an informative diagnostic of spin--orbit misalignment} in binary black hole mergers.
In particular, we considered the discrimination information between the isotropic- and aligned-spins hypotheses (as simplistic representatives of dynamical and isolated formation), which is given at the individual event level by the log Bayes factor and at the aggregate level by the Kullback--Leibler divergence.
We restricted these information measures to one spin-misalignment parameter at a time by approximating the Bayes factor through the Savage--Dickey ratio, and ascertained that $\cos{\theta_{LS}}$ is more informative than the effective precession spin $\chi_{\rm p}$, the opening angle $\cos\beta = \bm{\hat L}\cdot \bm{\hat J}$, or the primary spin tilt $\cos\theta_1 = \bm{\hat L} \cdot \bm{\hat \chi}_1$.
In addition to having informative posteriors, $\cos{\theta_{LS}}$ has the additional advantages compared to $\chip$ or $\cos\beta$ that it distinguishes between spin--orbit alignment and anti-alignment, and it has a well described and behaved prior distribution under both the aligned- and isotropic-spins hypotheses.
A limitation is that $\cos{\theta_{LS}}$ becomes ill defined for zero spin.
Like the other spin-misalignment parameters, $\theta_{LS}$ evolves on the precession timescale \cite{Apostolatos+94}, and it would be interesting to study its dynamics and interplay with spin--orbit resonances \cite{Gerosa2014}.

We analyzed the posterior distributions of events in GWTC-5.0 and \textbf{identified events that are strong candidates for spin--orbit misalignment using $\cos{\theta_{LS}}$}.
The parameter $\cos{\theta_{LS}}$ identifies events that can be diagnosed as precessing from other parameters including events GW241127\_061008 \citep{GWTC5}, GW241110\_124123 \citep{Abac2025b}, and GW241225\_082815 \citep{GWTC5}, which have claims of spin--orbit misalignment in existing literature.
Of these, GW241127\_061008 makes a very compelling case for a second-generation system, i.e.\ involving the remnant from a previous merger that was retained in its host stellar cluster.
This system has a large, tilted primary spin consistent with $\chi_1 \approx 0.7$, unequal masses consistent with a ratio of $q \approx 0.5$, and a large primary mass near or in the pair-instability-supernova gap (Fig.~\ref{fig:gw241127_cornerplot}).
We also pointed out additional less discussed events where spin--orbit alignment is ruled out. In several cases (such as GW231226\-\_101520, GW240920\-\_124024, GW240615\-\_113620, GW250114\-\_082203, GW200225\-\_060421, GW230723\_101834, GW150914\-\_090545, GW241230\-\_084504, GW231206\_233901, and GW250109\-\_010541) the black holes could possibly be nonspinning, and in others they must be spinning and misaligned (e.g., GW240920\-\_073424 and GW191109\-\_010717).
Through the implementation of $\cos{\theta_{LS}}$ we are able to provide a more complete analysis of the spin--orbit alignment of individual binary black hole mergers.

Finally, we demonstrated that \textbf{the astrophysical population of binary black hole systems requires spin--orbit misalignment}, in agreement with previous works \cite{Callister2022, Callister2024, gwtc5population}.
Simple population models with both highly constrained and free hyperparameters point to the existence of spin--orbit misalignment in the $\ctls$ distribution, and all models find that fewer systems have $\ctls<0$ than $\ctls>0$, disfavoring an entirely isotropic distribution.
All the models we considered find that $\sim 10$--$40 \%$ of binary black hole systems have negligible spin ($\chi \lesssim 0.05$). 
Although our most flexible model (\texttt{Gaussian + isotropic}) admits a peak at $\ctls \approx 0.2$, it is not required by the data.
This may be a similar phenomenon to the spurious features observed by \citet{VitaleMould2025} in the inferred $\cos\theta_{1,2}$ distributions using catalogs with $\sim 300$ events. Since $\ctls$ is on average as well as or better constrained than $\cos\theta_{1,2}$, one might expect that fewer events would be needed to better constrain its true distribution, giving a more encouraging outlook than \citet{VitaleMould2025} found; quantifying this would be an interesting avenue for future work.
Finally, recent population studies have identified multiple mass-dependent spin subpopulations \citep{gwtc5population, Banagiri2026, Ray2026, Ray2026a, Cheng2026, Guttman2026, Hussain2026, AlvarezLopez2026, Rinaldi2026, Plunkett2026, Zeeshan2026}.
Although we do not model such subpopulations in this work, investigating the $\ctls$ distribution across them could provide further insight into their underlying formation mechanisms.

\section*{Data Availability}
The code used to test the parameters, analyze the LVK events, and generate figures, along with additional figures for all events in the GWTC-5.0 catalog have been made publicly available at \url{https://github.com/charlie-gibson/cos_theta_ls}.

\begin{acknowledgments}

We thank Maia Martinez and Daniel Holz for helpful discussions.
This work was supported by the National Science Foundation Research Experience for Undergraduates (NSF REU) program, the LIGO Laboratory Summer Undergraduate Research Fellowship program (NSF LIGO), and the California Institute of Technology Student-Faculty Programs.
IA acknowledges support from NSF grants
2012086 and 2309360.
JR acknowledges support from the Sherman Fairchild Foundation, from the Simons Collaboration on Black Holes and Strong Gravity through grant SFI-MPS-BH-00012593-07, from the Kavli Institute for Cosmological Physics through an endowment from the Kavli Foundation and its founder Fred Kavli, and from the Jonathan M.\ Nelson Center for Collaborative Research.

This research has made use of data or software obtained from the Gravitational Wave Open Science Center (gwosc.org), a service of the LIGO Scientific Collaboration, the Virgo Collaboration, and KAGRA. This material is based upon work supported by NSF's LIGO Laboratory which is a major facility fully funded by the National Science Foundation, as well as the Science and Technology Facilities Council (STFC) of the United Kingdom, the Max-Planck-Society (MPS), and the State of Niedersachsen/Germany for support of the construction of Advanced LIGO and construction and operation of the GEO600 detector. Additional support for Advanced LIGO was provided by the Australian Research Council. Virgo is funded, through the European Gravitational Observatory (EGO), by the French Centre National de Recherche Scientifique (CNRS), the Italian Istituto Nazionale di Fisica Nucleare (INFN) and the Dutch Nikhef, with contributions by institutions from Belgium, Germany, Greece, Hungary, Ireland, Japan, Monaco, Poland, Portugal, Spain. KAGRA is supported by Ministry of Education, Culture, Sports, Science and Technology (MEXT), Japan Society for the Promotion of Science (JSPS) in Japan; National Research Foundation (NRF) and Ministry of Science and ICT (MSIT) in Korea; Academia Sinica (AS) and National Science and Technology Council (NSTC) in Taiwan.
\end{acknowledgments}

\bibliographystyle{apsrev4-2}
\bibliography{apssamp}

\end{document}